\documentclass[usenatbib]{mn2e}
\usepackage{graphicx}

\title[Southern SU UMa-Type Dwarf Novae]
{Photometric study of southern SU~UMa-type dwarf novae and candidates
   -- III: NSV~10934, MM~Sco, AB~Nor, CAL~86}

\author[T. Kato et al.]
{\parbox[t]{\textwidth}{
       Taichi Kato$^1$,
       Peter Nelson$^2$, 
       Chris Stockdale$^3$,
       Berto Monard$^4$, \\
       Tom Richards$^5$,
       Rod Stubbings$^6$, 
       Hitoshi Yamaoka$^7$,
       Bernard Heathcote$^8$, \\
       Roland Santallo$^9$ \\
} \\
       $^1$ Department of Astronomy, Faculty of Science,
       Kyoto University, Sakyo-ku, Kyoto 606-8502 Japan \\
       $^2$ RMB 2493, Ellinbank 3820, Australia \\
       $^3$ Hazelwood Observatory, RMB 4036 Matta Drive,
       Hazelwood South, Victoria 3840, Australia \\
       $^4$ Bronberg Observatory, PO Box 11426, Tiegerpoort 0056,
       South Africa \\
       $^5$ Woodridge Observatory, 8 Diosma Rd, Eltham, Vic 3095, Australia \\
       $^6$ 19 Greenland Drive, Drouin 3818, Victoria, Australia \\
       $^7$ Faculty of Science, Kyushu University, Fukuoka 810-8560,
       Japan \\
       $^8$ Tardis Astronomical Observatory, Mia Mia, Victoria, Australia \\
       $^9$ Southern Stars Observatory, Po Box 60972, 98702 FAAA TAHITI,
       French Polynesia \\
}

\date{Accepted.
      Received;
      in original form}

\pagerange{\pageref{firstpage}--\pageref{lastpage}}
\pubyear{2003}

\begin{document}

\maketitle

\label{firstpage}

\begin{abstract}
We photometrically observed four southern dwarf novae in outburst
(NSV 10934, MM Sco, AB Nor and CAL 86).  NSV 10934 was confirmed to be
an SU UMa-type dwarf nova with a mean superhump period of
0.07478(1) d.  This star also showed transient appearance of
quasi-periodic oscillations (QPOs) during the final growing stage of
the superhumps.
Combined with the recent theoretical interpretation and with the rather
unusual rapid terminal fading of normal outbursts, NSV 10934 may be
a candidate intermediate polar showing SU UMa-type properties.
The mean superhump periods of MM Sco and AB Nor were determined to be
0.06136(4) d and 0.08438(2) d, respectively.  We suggest that AB Nor
belongs to a rather rare class of long-period SU UMa-type dwarf novae
with low mass-transfer rates.  We also observed an outburst of the
suspected SU UMa-type dwarf nova CAL 86.  We identified this outburst
as a normal outburst and determined the mean decline
rate of 1.1 mag d$^{-1}$.
\end{abstract}


\section{Introduction}

   Cataclysmic variables (CVs) are close binary systems consisting of
a white dwarf and a red-dwarf secondary transferring matter via
Roche-lobe overflow.  SU UMa-type dwarf novae comprise an important
subgroup of CVs, which is characterized by the presence of superoutbursts
and superhumps.  The superhumps and superoutbursts are now widely
believed to be a result of the combination of two types of
disk-instabilities (thermal and tidal instabilities), which have provided
a laboratory to understand the basic astrophysical processes, such as
the origin of viscosity and resonant actions on a fluid disk in
close binaries (see a review by \citet{osa96review};
see also \citealt{ogi02tidal} for recent theoretical development).
We, the VSNET Collaboration \citep{VSNET},\footnote{
http://www.kusastro.kyoto-u.ac.jp/vsnet/.
} have been studying the properties of (mostly new) southern
SU UMa-type dwarf novae, candidates, and related systems with
a perspective described in \citet{kat03v877arakktelpucma}.
In this paper, we report on the
detection of superhumps in three systems, and also report on photometric
observations of an SU UMa-type candidate which underwent a likely
normal outburst.

\section{CCD Observation}

   The observers, equipment and reduction software are summarized in
Table \ref{tab:equipment}.  All observers performed aperture photometry,
and the magnitudes were determined relative to a nearby comparison star,
which was confirmed to be constant during the observation by
a comparison with a check star.
The observations used unfiltered CCD systems having a response
close to Kron-Cousins $R_{\rm c}$ band for outbursting dwarf novae.
The errors of single measurements are typically
less than 0.01--0.03 mag unless otherwise specified.
The observers abbreviations will be used in ``Obs'' field in the
later observing logs.

\begin{table}
\caption{Observers and Equipment.} \label{tab:equipment}
\begin{center}
\begin{tabular}{cccc}
\hline\hline
Observer (Abbr.) & Telescope$^a$ &  CCD  & Software \\
\hline
Heathcote (H)  & 35.5-cm SCT & Audine & AIP4Win \\
Nelson    (N)  & 32-cm reflector & ST-8E & AIP4Win \\
Monard    (M)  & 30-cm SCT & ST-7E & AIP4Win \\
Richards  (R)  & 18-cm refractor & ST-7E & AIP4Win \\
Santallo  (Sa) & 20-cm SCT & ST-7E & AIP4Win \\
Stockdale (St) & 28-cm SCT & Meade 416XTE & MaxIm$^b$ \\
               &           &              & AIP4Win \\
\hline
 \multicolumn{4}{l}{$^a$ SCT = Schmidt-Cassegrain telescope.} \\
 \multicolumn{4}{l}{$^b$ Used for NSV 10934.} \\
\end{tabular}
\end{center}
\end{table}

   Barycentric corrections to the observed times were applied before the
following analysis.

\section{NSV 10934}

\subsection{Introduction}

   NSV 10934 was originally discovered as a large-amplitude suspected
variable star of unknown classification.  \citet{kat02gzcncnsv10934}
noticed the identification with a bright ROSAT source
(1RXS J184050.3$-$834305), and suggested that the object is a cataclysmic
variable.  \citet{kat02gzcncnsv10934} indeed detected multiple outbursts.
These outbursts generally bore resemblance to dwarf nova outbursts, but
are unusual in the rapid decline during the terminal stages of these
outbursts.  From these findings, \citet{kat02gzcncnsv10934} suggested that
NSV 10934 may be an analogous object to the intermediate polar (IP),
HT Cam, which shows brief dwarf nova-like outbursts
\citep{ish02htcam,kem02htcam}.  \citet{kat02gzcncnsv10934} also predicted
that the orbital period of NSV 10934 would be slightly longer than that
of HT Cam (86 min), if NSV 10934 indeed turns out to be an HT Cam-like
object.

\begin{figure}
  \includegraphics[angle=0,width=8.8cm]{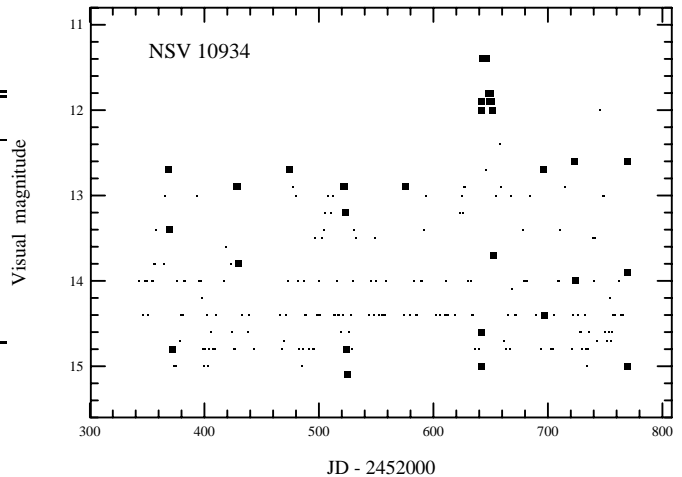}
  \caption{Long-term visual light curve of NSV 10934.  In addition to
  short, normal outbursts recurring with time-scales of 40--60 d,
  there is a long, bright superoutburst around JD 2452643.  The enlarged
  CCD light curve of this superoutburst is shown in
  Figure \ref{fig:nsvlc}.}
  \label{fig:nsvout}
\end{figure}

   Since then, NSV 10934 has been monitored by one of the authors
(Rod Stubbings), and it has been established, by the end of 2002, that
the object shows short outbursts, as described in
\citet{kat02gzcncnsv10934}, at rather regular intervals
of 40--60 d.  Following a call for observing campaign in 2002 December
(vsnet-campaign-dn 3141\footnote{
http://www.kusastro.kyoto-u.ac.jp/vsnet/Mail/\\campaign-dn3000/msg00141.html.
}), the object went into a superoutburst, which will be described
in the next subsection.  A recent long-term visual light curve is
shown in Figure \ref{fig:nsvout}.

\subsection{2003 January Outburst}

   The outburst was first detected on 2003 January 2.480 UT at a visual
magnitude of 15.0 by Rod Stubbings.  On January 2.767 UT, the object
was observed to further brighten to 11.9 mag
(vsnet-alert 7601\footnote{
http://www.kusastro.kyoto-u.ac.jp/vsnet/Mail/alert7000/\\msg00601.html.
}).  A time-resolved CCD photometric campaign started on the next night
of this detection.  The object was still rising in brightness.
As described later, this outburst was confirmed to be a superoutburst
by the detection of secure superhumps.
The log of observation is summarized in Table \ref{tab:nsvlog}.

\begin{figure}
  \includegraphics[angle=0,width=8.8cm]{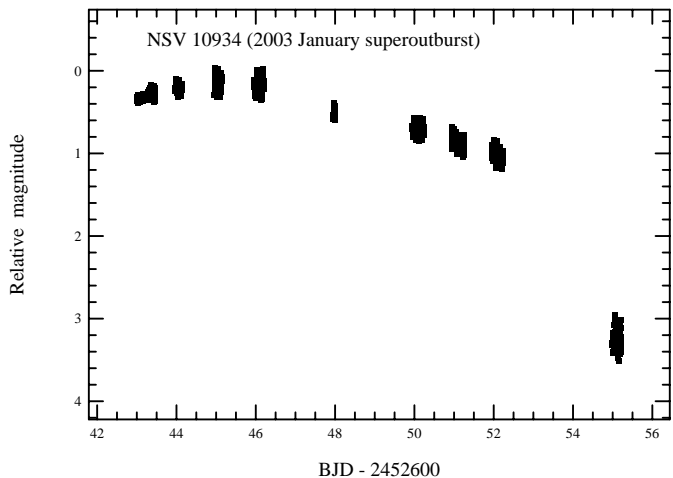}
  \caption{The 2003 January superoutburst of NSV 10934.  The CCD magnitudes
  (close to $R_{\rm c}$) are given relative to GSC 9523.351
  ($R_{\rm c} \sim$11.8).}
  \label{fig:nsvlc}
\end{figure}

   Figure \ref{fig:nsvlc} shows the entire light curve of this superoutburst
drawn from CCD observations.  The CCD magnitudes (system close to
$R_{\rm c}$)
are given relative to GSC 9523.351 ($R_{\rm c} \sim$11.8).  The zero point
is adjusted to the most comprehensive Peter Nelson's observations.

\begin{table}
\caption{Journal of the 2003 CCD photometry of NSV 10934.}\label{tab:nsvlog}
\begin{center}
\begin{tabular}{crccrc}
\hline\hline
\multicolumn{2}{c}{2003 Date}& Start--End$^a$ & Exp(s) & $N$
        & Obs \\
\hline
Jan.  &  3 & 52643.014--52643.159 &  15  & 192 & St \\
      &  3 & 52643.271--52643.467 &  20  & 626 & M \\
      &  4 & 52643.970--52644.131 &  15  & 276 & St \\
      &  5 & 52644.958--52645.137 &  15  & 281 & St \\
      &  5 & 52644.962--52645.080 &  15  & 334 & N \\
      &  6 & 52645.944--52646.134 &  15  & 256 & St \\
      &  6 & 52645.966--52646.185 &  10  & 685 & N \\
      &  6 & 52645.978--52646.205 &  30  & 443 & R \\
      &  8 & 52647.958--52648.010 &  20  & 130 & N \\
      & 10 & 52649.942--52650.125 &  15  & 292 & St \\
      & 10 & 52649.958--52650.241 &  40  & 492 & R \\
      & 10 & 52649.985--52650.235 &  15  & 628 & N \\
      & 11 & 52650.938--52651.125 &  15  & 214 & St \\
      & 11 & 52650.963--52651.263 &  40  & 492 & R \\
      & 12 & 52651.951--52652.123 &  15  & 335 & St \\
      & 12 & 52651.989--52652.233 &  20  & 525 & N \\
      & 12 & 52652.013--52652.246 &  40  & 400 & R \\
      & 15 & 52654.992--52655.219 & 180  &  98 & N \\
\hline
 \multicolumn{6}{l}{$^a$ BJD$-$2400000.} \\
\end{tabular}
\end{center}
\end{table}

\subsection{Superhumps}

\begin{figure}
  \includegraphics[angle=0,width=8.8cm,height=11cm]{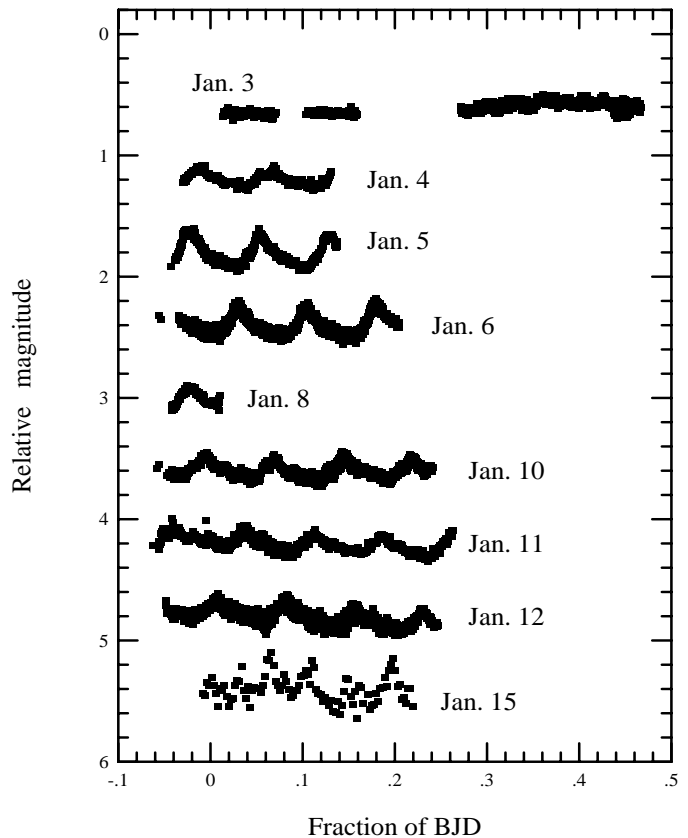}
  \caption{Nightly light curves of NSV 10934 during the 2003 January
  superoutburst.  The light curve on
  January 3 showed almost no feature of superhumps.  On January 4,
  superhumps started to grow.  The amplitudes of the superhumps reached
  a maximum around January 5--6.}
  \label{fig:nsvday}
\end{figure}

\begin{figure}
  \includegraphics[angle=0,width=8.8cm]{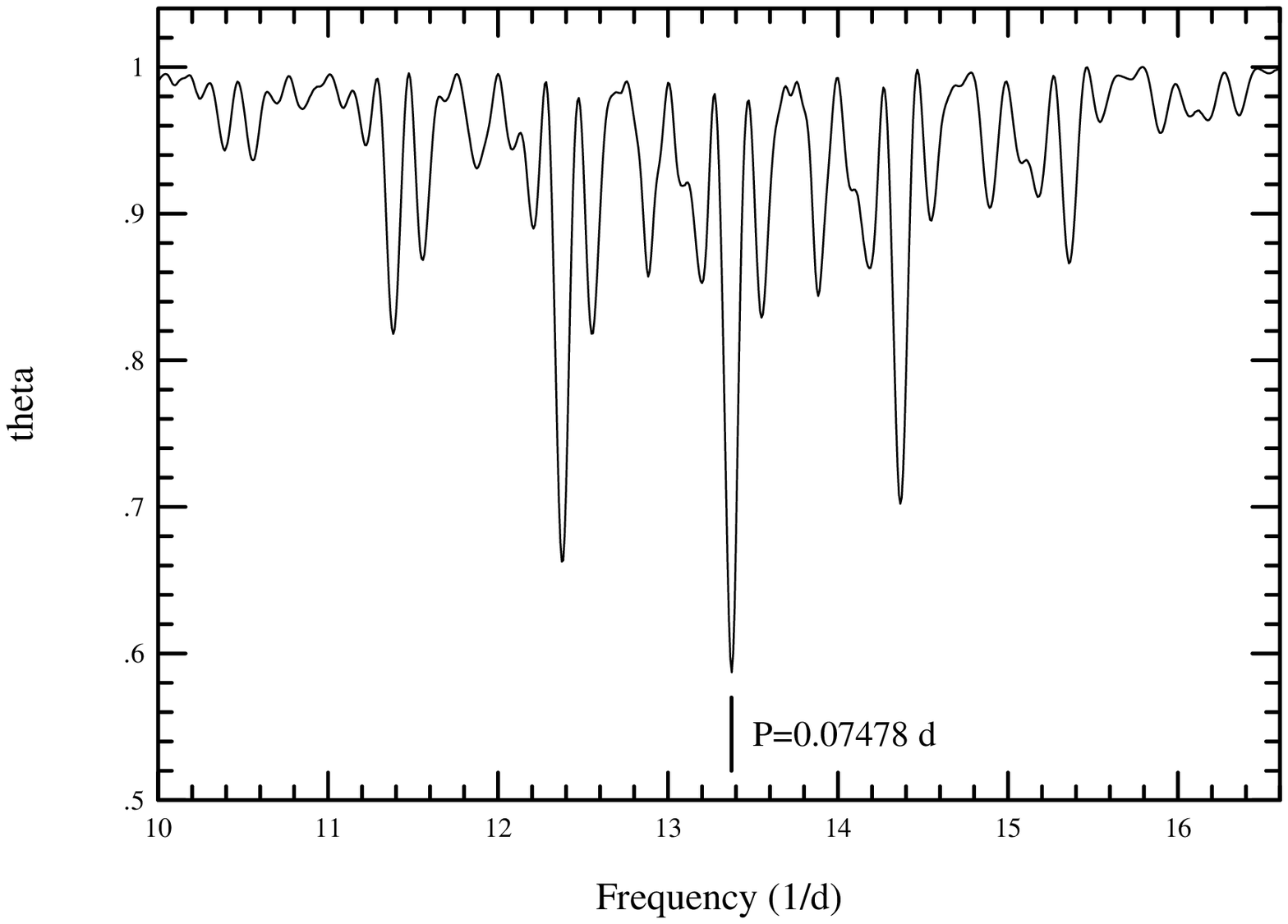}
  \caption{Period analysis of NSV 10934 (plateau stage: 2003 January 5--13).
  The strongest signal at a frequency
  of 13.373(1) d$^{-1}$ corresponds to the best superhump period
  of 0.07478(1) d.}
  \label{fig:nsvpdm}
\end{figure}

   Figure \ref{fig:nsvday} shows nightly light curves of NSV 10934
during the 2003 January superoutburst.  The light curve on January 3 showed
almost no feature of superhumps.  On January 4, superhumps started to
grow.  The amplitudes of the superhumps reached a maximum at around
January 5--6.  We first determined the mean superhump period using
Phase Dispersion Minimization (PDM: \citealt{PDM}), after removing the
linear decline trend during the plateau stage (January 5--13) of the
superoutburst.  The result is shown in Figure \ref{fig:nsvpdm}.
The strongest signal at a frequency of 13.373(1) d$^{-1}$ corresponds to
the best superhump period of 0.07478(1) d.  The selection of the
correct alias has been confirmed by independent analyses of continuous
nightly observations.

\begin{figure}
  \includegraphics[angle=0,width=8.8cm]{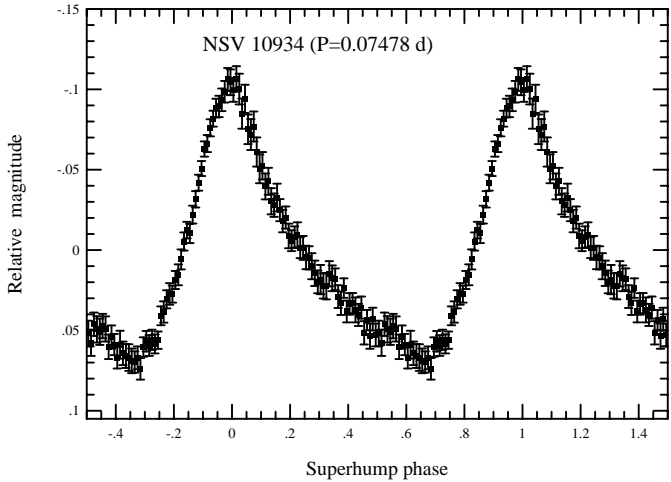}
  \caption{Mean superhump profile of NSV 10934.}
  \label{fig:nsvph}
\end{figure}

   Figure \ref{fig:nsvph} shows the mean superhump profile
(of the plateau phase of the superoutburst), phase-averaged with the period
of 0.07478 d.  The rapid rise and slower decline are
characteristic of SU UMa-type superhumps \citep{vog80suumastars,war85suuma}.

\subsection{Superhump period change}

   We extracted the maximum times of superhumps from the light curve by eye.
The averaged times of a few to several points close to the maxima were
used as representatives of the maximum times.  Thanks to the
high-precision data, the errors of the maximum times are usually less
than $\sim$0.001 d.  The resultant superhump maxima
are given in Table \ref{tab:nsvmax}.  The values are given to 0.0001 d in
order to avoid the loss of significant digits in a later analysis.
The cycle count ($E$) is defined as the cycle number since BJD 2452643.9874.
A linear regression to the observed superhump times gives the following
ephemeris (the errors correspond to 1$\sigma$ errors at $E$ = 48):

\begin{equation}
{\rm BJD (maximum)} = 2452644.0033(16) + 0.074851(38) E. \label{equ:nsvreg1}
\end{equation}

\begin{table}
\caption{Times of superhump maxima of NSV 10934.}\label{tab:nsvmax}
\begin{center}
\begin{tabular}{ccc}
\hline\hline
$E^a$  & BJD$-$2400000 & $O-C^b$ \\
\hline
  0 & 52643.9874 &  -0.0159 \\
  1 & 52644.0693 &  -0.0089 \\
 13 & 52644.9766 &   0.0002 \\
 14 & 52645.0537 &   0.0024 \\
 15 & 52645.1289 &   0.0028 \\
 27 & 52646.0301 &   0.0058 \\
 28 & 52646.1049 &   0.0057 \\
 29 & 52646.1791 &   0.0051 \\
 53 & 52647.9784 &   0.0080 \\
 80 & 52649.9951 &   0.0037 \\
 81 & 52650.0707 &   0.0045 \\
 82 & 52650.1470 &   0.0059 \\
 83 & 52650.2161 &   0.0002 \\
 94 & 52651.0398 &   0.0004 \\
107 & 52652.0082 &  -0.0042 \\
108 & 52652.0825 &  -0.0048 \\
109 & 52652.1569 &  -0.0052 \\
110 & 52652.2315 &  -0.0055 \\
\hline
 \multicolumn{3}{l}{$^a$ Cycle count since BJD 2452643.9874.} \\
 \multicolumn{3}{l}{$^b$ $O-C$ calculated against equation
                    \ref{equ:nsvreg1}.} \\
\end{tabular}
\end{center}
\end{table}

\begin{figure}
  \includegraphics[angle=0,width=8.8cm]{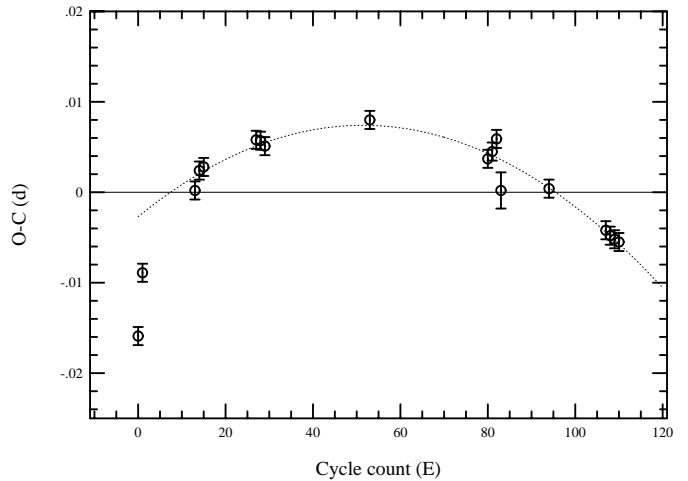}
  \caption{$O-C$ diagram of superhump maxima of NSV 10934.  The error
  bars correspond to the upper limits of the errors except for
  $E$ = 83, which has a larger error 0.002 d.
  The parabolic fit for 13$\leq E\leq$110 is shown with a dotted line.}
  \label{fig:nsvoc}
\end{figure}

   Figure \ref{fig:nsvoc} shows the ($O-C$)'s against the mean superhump
period (0.074851 d) from a linear regression (equation \ref{equ:nsvreg1}).
The diagram clearly shows the decrease in the superhump period throughout
the superoutburst plateau.  The superhump maxima during the plateau phase
(13$\leq E\leq$110) is well expressed by a quadratic term corresponding
to a period derivative of
$\dot{P}/P$ = $-$10.2$\pm$1.0 $\times$ 10$^{-5}$.  The earlier stage
($E<$13) shows a large deviation from this quadratic fit, which is
probably a result of the rapid evolution of superhumps at this
stage.  The observed decrease of the superhump period is one of the
largest among known SU UMa-type dwarf novae
(cf. \citealt{kat03v877arakktelpucma}).

\subsection{Super-QPOs?}\label{sec:nsvqpo}

\begin{figure}
  \includegraphics[angle=0,width=8.8cm]{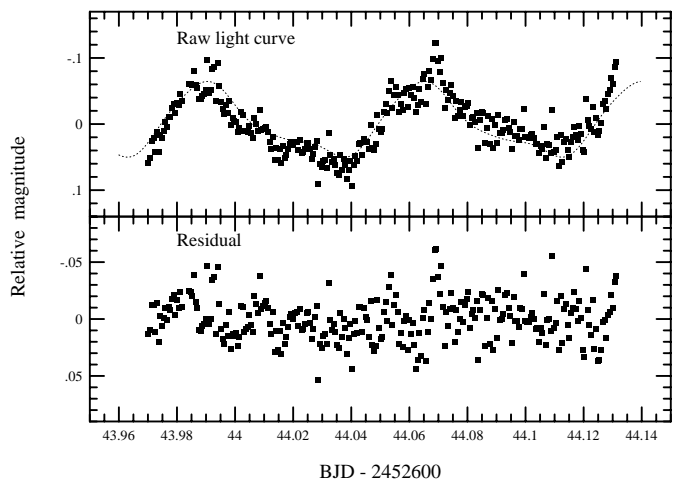}
  \caption{Enlarged light curve NSV 10934 on 2003 January 4.
  (Upper:) Raw data (the dotted line represents the best-fit superhump
  signal). (Lower:) Residual light curve subtracted for the best-fit
  mean superhump light curve.  Quasi-periodic oscillations with periods
  $\sim$0.015 d are present.}
  \label{fig:nsv4}
\end{figure}

\begin{figure}
  \includegraphics[angle=0,width=8.8cm]{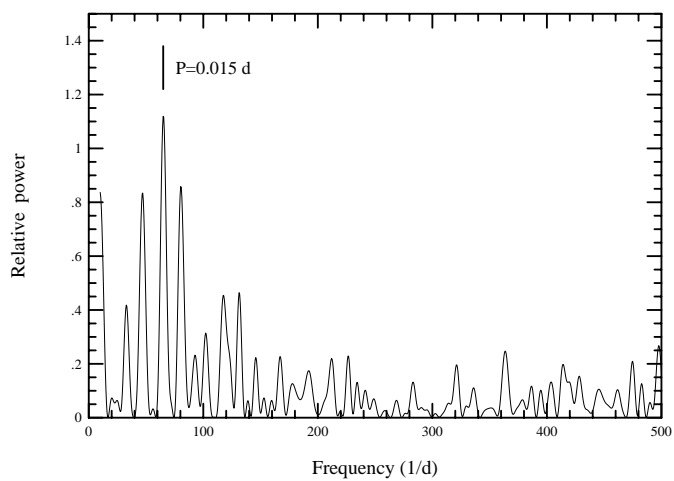}
  \caption{Power spectrum of the QPOs on 2003 January 4.  The strongest
  signal is at a frequency of 65 d$^{-1}$, corresponding to a period of
  0.015 d.}
  \label{fig:nsv4pow}
\end{figure}

   On January 4 (just after the initial growth time of the superhumps),
there was an indication of quasi-periodic oscillations (QPOs) superimposed
on superhumps (Figure \ref{fig:nsv4}).  The lower panel of Figure
\ref{fig:nsv4} is to better illustrate the QPO signal, by subtracting
the mean superhump profile by using a Fourier decomposition of
the superhump profile from these data up to the fourth harmonics.
Figure \ref{fig:nsv4pow} shows the power spectrum of the QPOs.
The strongest signal was found at a frequency of 65 d$^{-1}$,
corresponding to a period of 0.015 d.
No comparable QPOs were observed on preceding and following nights.
This transient appearance of the QPO signal is very reminiscent of
``super-QPOs" in some SU UMa-type dwarf novae, which only appear
during the early stage of the superhump evolution
\citep{kat92swumasuperQPO,kat02efpeg}.  \citet{war02DNO} suggested
that these super-QPOs would be a result of interaction between the
weak magnetism of the white dwarf and some kind of wave in the
inner accretion disk.
If this interpretation could apply to NSV 10934, the possible
intermediate polar-type interpretation of this object
\citep{kat02gzcncnsv10934} would be consistent with the present finding.

\subsection{NSV 10934 as an SU UMa-Type Dwarf Nova}

   Although the supercycle length of NSV 10934 has not yet been
established, the intervals between normal outbursts (40--60 d) are typical
values for an SU UMa-type dwarf nova in the intermediate activity class
\citep{vog93suuma}.  The apparent lack of terminal brightening during
the superoutburst plateau also fits the general properties of SU UMa-type
dwarf nova with this superhump period \citep{kat03hodel}.
However, the presence of terminal rapid declines
during normal outbursts \citep{kat02gzcncnsv10934} is rather
unusual, because the contribution from quiescent luminosity usually
works to slow down the decline rate near the terminal stage of
such outbursts (e.g. \citealt{vanpar94suumayzcnc}).
It may be that the inner accretion
disk is truncated by the weak magnetic field of the white dwarf
to produce such rapid terminal declines (e.g. \citealt{ish02htcam}),
while the field strength is not strong enough to moderate dwarf nova-type
outburst properties (e.g. \citealt{ang89DNoutburstmagnetic}).  As stated
in section \ref{sec:nsvqpo}, the supposed presence of a weak magnetic
field would naturally explain the appearance of super-QPOs at the
same time.  A search for a coherent signal in X-ray, ultraviolet,
and optical wavelengths is encouraged.  The only other SU UMa-type
dwarf nova which was proposed to be an IP is VZ Pyx
\citep{rem94v1028tauvzpyx,alv95vzpyxIUE,kat97vzpyx,kiy99vzpyx}.
Although this object was originally proposed to be an IP, the modern
observational evidence is rather against this classification
(cf. \citealt{kat97vzpyx,war03DNO3}).\footnote{
  The well-known SU UMa-type dwarf nova SW UMa is also suspected to be
  an IP \citep{sha86swumaXray,rob87swumaQPO,
  szk88swumaEXOSATIUE,ros94v426ophswumav348pup},
  although the outburst parameters of SW UMa are rather unusual
  among SU UMa-type dwarf novae
  \citep{sha87swuma,how95TOAD,how95swumabcumatvcrv}.  It is possible
  that this possible IP nature may be responsible for the most striking
  appearance of super-QPOs \citep{kat92swumasuperQPO}.
}

\section{MM S\lowercase{co}}

\subsection{Introduction}

   MM Sco was discovered as a dwarf nova on Harvard plates
(cf. \citealt{GlasbyDNbook,wal58CVchart}).
\citet{pet56uvper} suggested, from the apparently long outburst interval
($\geq$500 d), that MM Sco may a similar object to UV Per, which
is currently known as an SU UMa-type dwarf nova with long supercycles
(\citealt{uda92uvper,kat90uvper}; see also \citealt{kat01hvvir} for a discussion
on the relation to WZ Sge-type dwarf novae).
This cycle length was adopted in \citet{GCVS3}.
However, F. M. Bateson suggested that the mean outburst cycle length
($\sim$28 d) is much shorter than what has been believed, and observed
maxima were fainter than the originally reported magnitude (see the
description in \citet{vog83DNUBV}; this period was adopted in \citet{GCVS};
for a more recent reference, see \citealt{bat97mmsco}).  The reported
outburst characteristics of MM Sco was thus rather controversial.\footnote{
  In a most recent publication, \citet{mas03faintCV} listed MM Sco
  as a candidate SU UMa-type dwarf nova, though they reported that
  no superhumps have yet been observed.
}

\subsection{2002 September Outburst}

   MM Sco has been monitored by the VSNET members since the outburst in
1997 (cf. vsnet-alert 946\footnote{
http://www.kusastro.kyoto-u.ac.jp/vsnet/Mail/vsnet-alert/\\msg00946.html.
}) because of its apparently low frequency of outbursts, which is
a rather commonly met signature of SU UMa-type dwarf novae.

   The 2002 September outburst was detected by Rod Stubbings on 2002
September 5.428 UT at a visual magnitude of 14.0
(vsnet-outburst 4485\footnote{
http://www.kusastro.kyoto-u.ac.jp/vsnet/Mail/outburst4000/\\msg00485.html.
}).  The object further brightened to a magnitude of 13.4 next night,
indicating that the present outburst may be a long, bright outburst.
We carried out time-resolved CCD photometry upon this information.
The log of observation is summarized in Table \ref{tab:mmlog}.

\begin{table}
\caption{Journal of the 2002 CCD photometry of MM Sco.}\label{tab:mmlog}
\begin{center}
\begin{tabular}{crccrc}
\hline\hline
\multicolumn{2}{c}{2002 Date}& Start--End$^a$ & Exp(s) & $N$
        & Obs \\
\hline
Sept. & 10 & 52528.221--52528.390 &  45  & 214 & M \\
      & 11 & 52529.200--52529.436 &  45  & 327 & M \\
      & 12 & 52529.715--52529.828 &  20  & 328 & Sa \\
\hline
 \multicolumn{6}{l}{$^a$ BJD$-$2400000.} \\
\end{tabular}
\end{center}
\end{table}

\begin{figure}
  \includegraphics[angle=0,width=8.8cm,height=11cm]{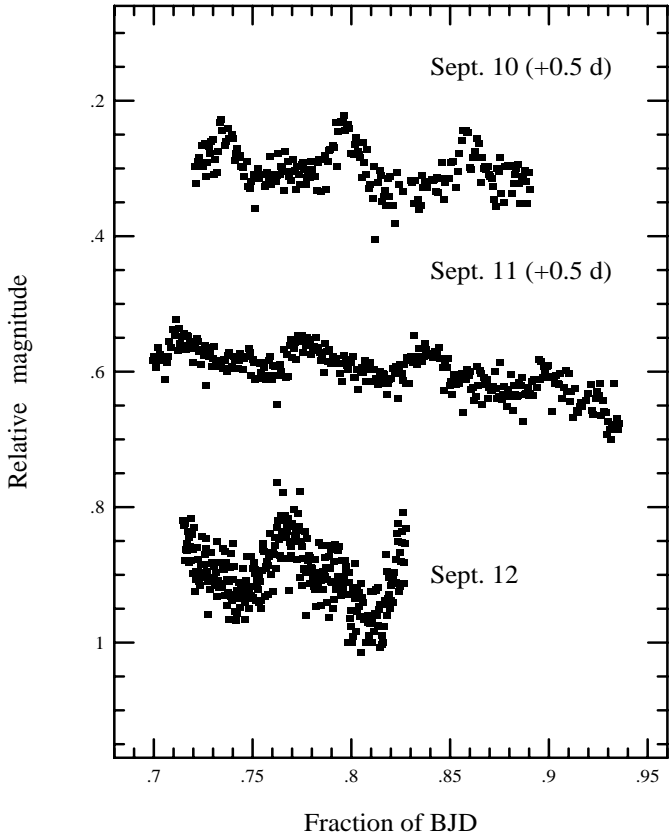}
  \caption{Nightly light curves of MM Sco.  Superhumps were clearly
  visible on all nights.}
  \label{fig:mmday}
\end{figure}

\begin{figure}
  \includegraphics[angle=0,width=8.8cm]{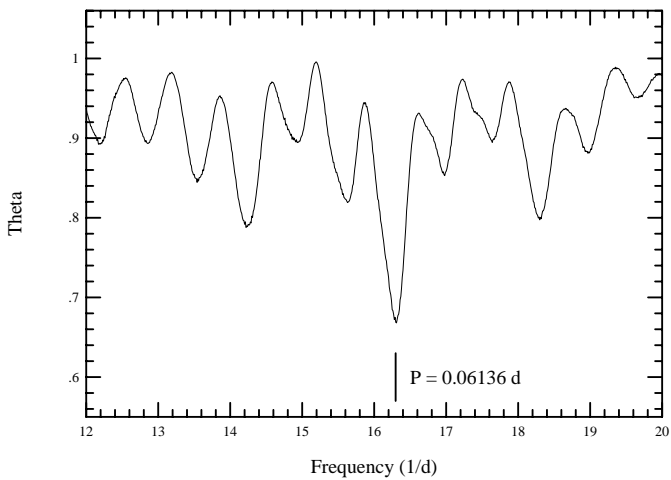}
  \caption{Period analysis of MM Sco.  The strongest signal at a frequency
  of 16.298(11) d$^{-1}$ corresponds to the best superhump period
  of 0.06136(4) d.}
  \label{fig:mmpdm}
\end{figure}

   Figure \ref{fig:mmday} shows the nightly light curves of MM Sco.
Superhumps were clearly visible on all nights; this confirms the
SU UMa-type nature of MM Sco.  Figure \ref{fig:mmpdm} shows the result
of a period analysis using PDM applied to the entire data set after
removing the nightly linear decline trends.
The strongest signal at a frequency
of 16.298(11) d$^{-1}$ corresponds to the best superhump period
of 0.06136(4) d.

\begin{figure}
  \includegraphics[angle=0,width=8.8cm]{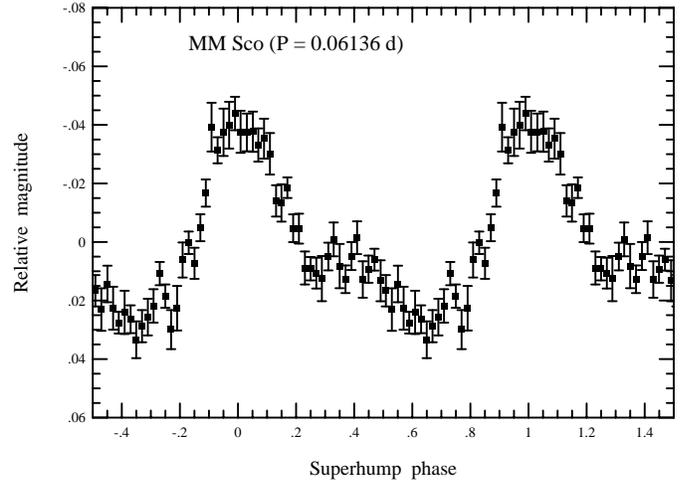}
  \caption{Mean superhump profile of MM Sco.}
  \label{fig:mmph}
\end{figure}

   Figure \ref{fig:mmph} shows the mean superhump profile phase-averaged
with the period of 0.06136 d.  The rapid rise and slower decline are
characteristic of SU UMa-type superhumps.
We did not attempt to determine a period derivative 
(e.g. \citealt{kat03v877arakktelpucma,kat03bfara}) because of the
short baseline of the observation.
The less sharp appearance of the superhump maximum, compared to
other fully grown superhumps of SU UMa-type dwarf novae (e.g.
\citet{har95cyuma}), was probably because the observation started
$\sim$5 d after the start of the superoutburst.  The superhumps may have
entered its decaying phase at the time of our observation.  Further
detailed observations of the full evolutionary course of the superhumps,
as well as determination of the orbital period,
are strongly encouraged to fully understand the behavior of
superhumps in this system.

\subsection{Astrometry and Quiescent Counterpart}

\begin{figure}
  \begin{center}
  \includegraphics[angle=0,width=8cm]{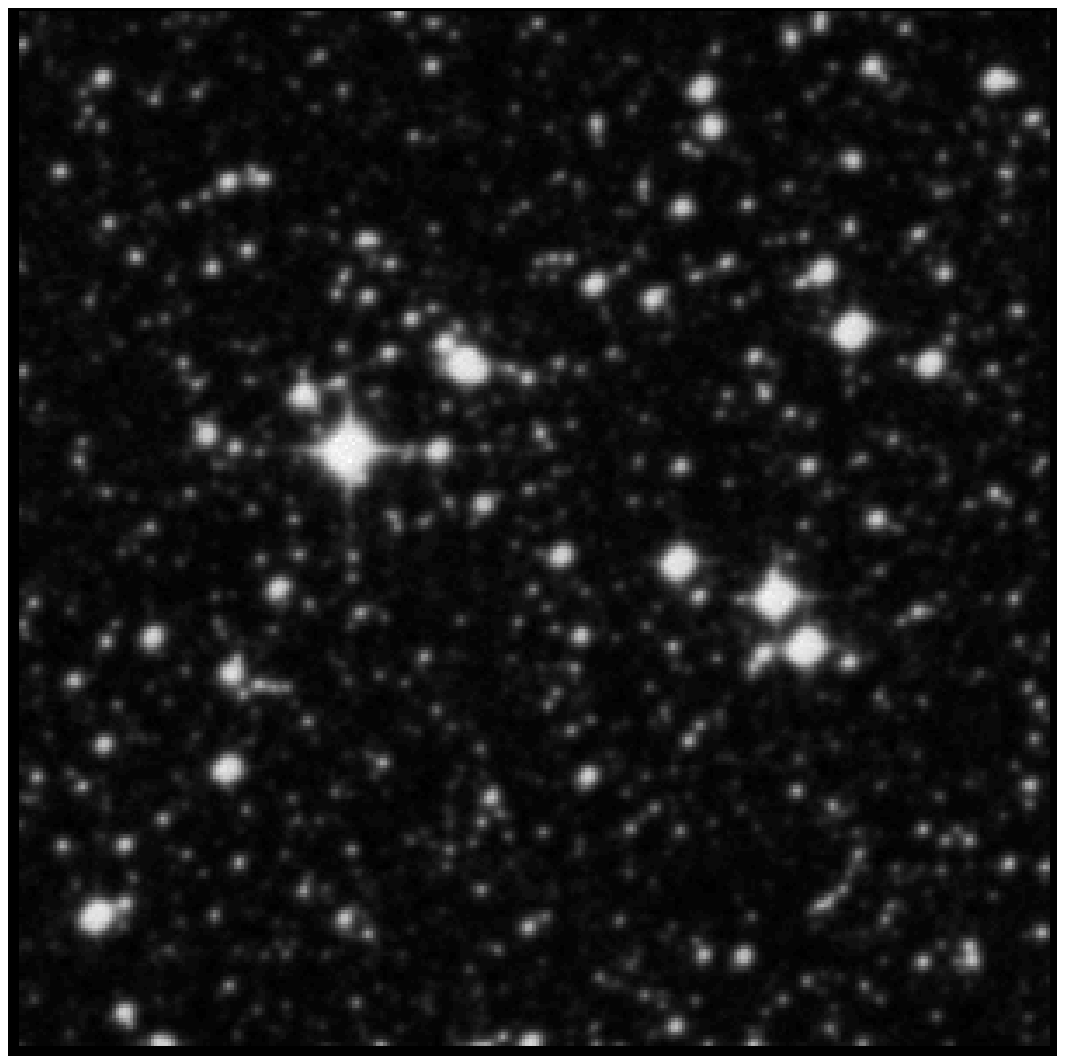} \\
  \vskip 1mm
  \includegraphics[angle=0,width=8cm]{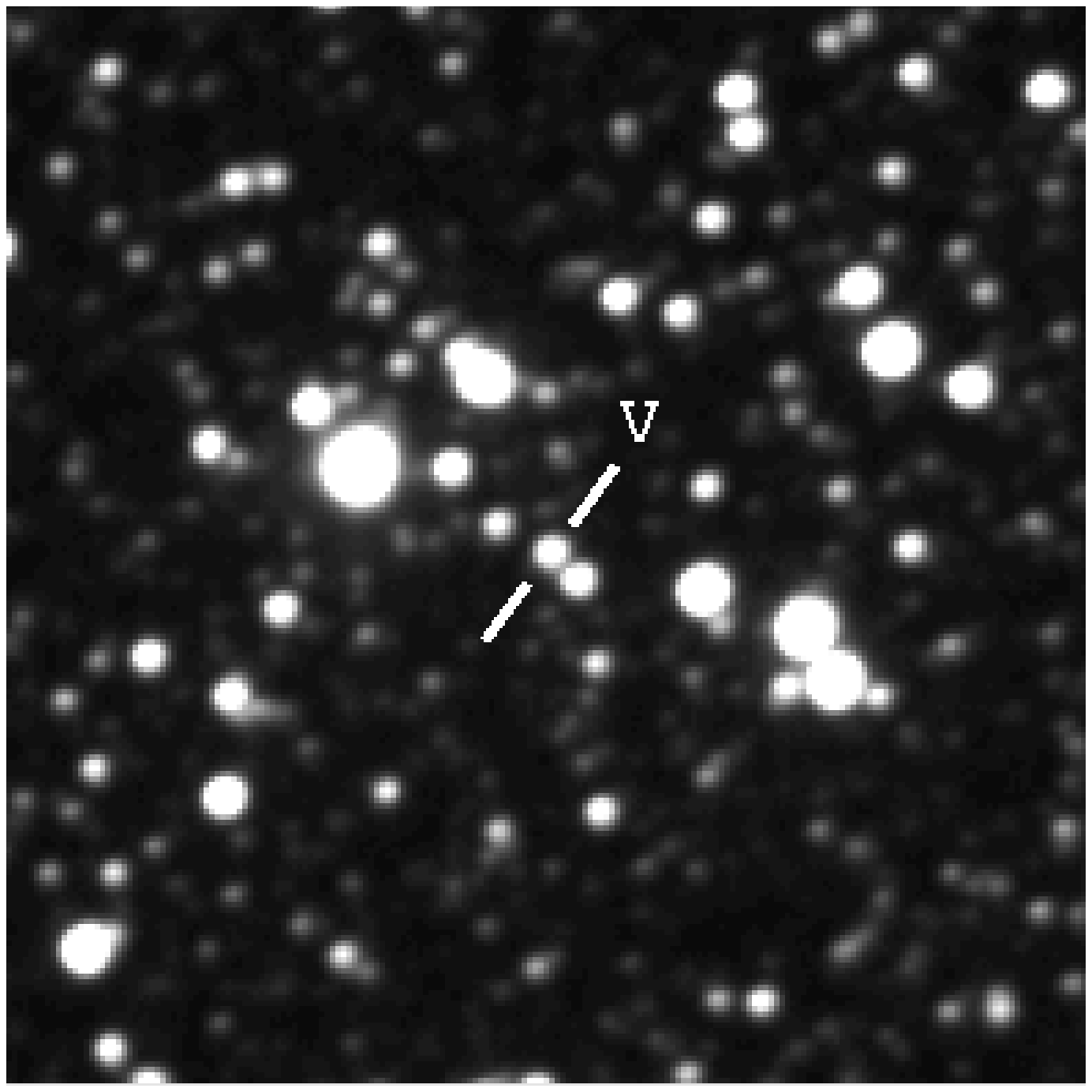}
  \end{center}
  \caption{Identification of MM Sco.  Up is north, left is east, 5
  minutes square.  (Upper) In quiescence, reproduced from the DSS 2
  red image.  (Lower) In outburst, taken on 2002 Sept. 10.77 UT by
  B. Monard. V = MM Sco.
  }
  \label{fig:mmscoid}
\end{figure}

  We measured the position of the outbursting object with respect to the
UCAC1 reference frame and yielded R. A. = 17$^{\rm h}$ 30$^{\rm m}$
45$^{\rm s}$.254, Decl. = $-$42$^\circ$ 11$'$ 42$''$.69 (J2000.0)
(Fig. \ref{fig:mmscoid}).
\citet{vog82atlas} exactly points the object at this position.
On the other hand, the DSS 2 images of this region show an only faint
object, implying that MM Sco was accidentally caught in a slightly
brightened state in \citet{vog82atlas}.  The DSS 2 star apparently moved
in the west-southwest direction between two plates
($I$-band, epch 1980.478 and $R$-band, epoch 1997.249).
This star is likely the object marked on Downes' online atlas.\footnote{
  http://icarus.stsci.edu/$^{\sim}$downes/cvcat.
}
There is a possibility that MM Sco in true quiescence is fainter than
the limit of the DSS 2 images and that the apparently moving object is
an unrelated star.  If it is the case, the outburst amplitude could
exceed 6 mag.  Definite quiescent identification and precise amplitude
measurement should await deep direct imaging with higher spatial
resolution.

\subsection{MM Sco as an SU UMa-Type Dwarf Nova}

\begin{figure*}
  \includegraphics[angle=0,width=18cm]{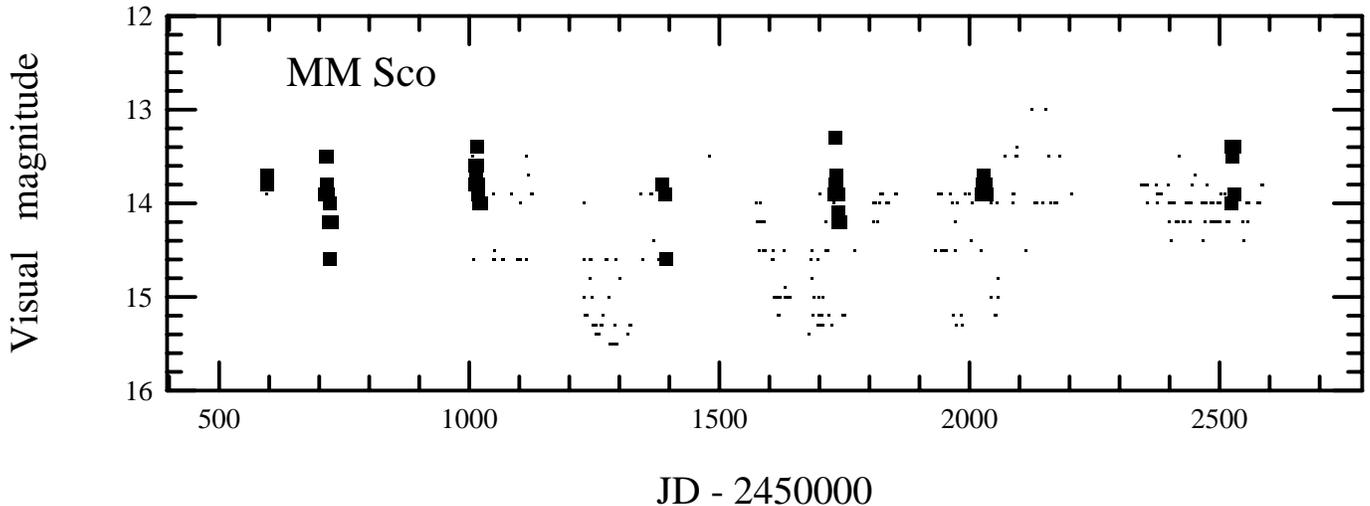}
  \caption{Long-term visual light curve of MM Sco.  Large and small dot
  represent positive detections and upper limit observations, respectively.
  Outbursts other than on JD 2450596 are superoutbursts.}
  \label{fig:mmlong}
\end{figure*}

   Figure \ref{fig:mmlong} shows the long-term visual light curve of MM Sco.
Table \ref{tab:mmout} lists the observed outbursts.
Six well-defined superoutbursts (JD 2450712, 2451010, 2451385, 2451729,
2452025, and 2452523) with durations longer than 8 d are unambiguously
identified.  The supercycle lengths are thus in the range of 298--497 d.
There does not seem to be a fixed supercycle length as recorded in
KK Tel \citep{kat03v877arakktelpucma}.

   In spite of the relatively bright superoutburst magnitudes
(usually 13.3--13.8), very few normal outbursts have been detected,
which could have easily reached detectable magnitudes.
Although the small number of observations makes it difficult to draw
a firm conclusion on the type of outburst, the outburst on JD 2450596 was
the only candidate normal outburst since 1997.  Such a low number ratio
normal outbursts over superoutbursts is exceptional
(cf. \citealt{war95suuma,nog97sxlmi}).  In combination with the relatively
short superhump period (0.06146 d), the initially proposed analogy
\citep{pet56uvper} with UV Per ($P_{\rm SH}$ = 0.06641 d) looks likes
to be more strengthened.
The shortest intervals (28 d) of outbursts may have corresponded to
a precursor outburst or a rebrightening phenomenon, both of which are
relatively commonly observed in SU UMa-type dwarf novae with short
superhump periods and less frequent normal outbursts
\citep{lem93tleo,pat93vyaqr,how95swumabcumatvcrv,kat97tleo,
nog98swuma,bab00v1028cyg,kat01wxcet,ish01rzleo}.

   This finding makes a contrast to what was originally suggested by
F. M. Bateson \citep{vog83DNUBV}.  It may be either possible that the
finding by F. M. Bateson did not correctly describe the outburst behavior
of this object due to the lack of appropriate information at that time,
or that the outburst characteristics
exhibited a long-term variation.  Since some SU UMa-type dwarf novae
are known to show dramatic long-term variation, particularly in the
number of normal outbursts (e.g. V503 Cyg: \citet{kat02v503cyg};
DM Lyr: \citet{nog03dmlyr}; MN Dra = Var73 Dra: \citet{nog03var73dra}),
this possibility in MM Sco needs to be carefully checked by
future observations.

\begin{table}
\caption{List of Outbursts of MM Sco.}\label{tab:mmout}
\begin{center}
\begin{tabular}{ccccc}
\hline\hline
JD start$^a$ & JD end$^a$ & Max & Duration (d) & Type \\
\hline
50595.9 &   --    & 13.7 &  --  & normal? \\
50712.0 & 50724.9 & 13.5 & $>$13 & super \\
51010.9 & 51022.9 & 13.5 & $>$13 & super \\
51385.9 & 51393.9 & 13.8 & $>$8 & super \\
51729.9 & 51742.0 & 13.3 & 13   & super \\
52025.0 & 52033.3 & 13.8 & $>$8 & super \\
52522.9 & 52529.9 & 13.4 & $>$7 & super \\
\hline
 \multicolumn{5}{l}{$^a$ JD$-$2400000.} \\
\end{tabular}
\end{center}
\end{table}

\section{AB N\lowercase{or}}

\subsection{Introduction}

   AB Nor was discovered by \citet{swo30abnorbrlup} during the photographic
survey of the southern Milky Way.  Only little had been studied until
very recent years.  \citet{pet60DNe} simply provided a ``long?" recurrence
period in the table of dwarf nova candidates.
\citet{vog82atlas} proposed a quiescent counterpart
based on its blue color, but the direct attempt to identify the object
by recording an outburst was not successful.  The first outburst reported
to VSNET was in 1997 (section \ref{sec:abnorsu}); the object has been
regularly monitored since then.

   An outburst in 2000 April, detected by the Rod Stubbings, was most
unusual.  Five days after the initial brightness peak decayed, the
object sudden underwent a rebrightening (vsnet-alert 4574\footnote{
http://www.kusastro.kyoto-u.ac.jp/vsnet/Mail/alert4000/\\msg00574.html.
}).  Since such an early-stage rebrightening is usually associated with
a superoutburst triggered by an immediately preceding precursor
(\citealt{mar79superoutburst,war85suuma,kat97tleo}; see also the PU CMa
case in \citet{kat03v877arakktelpucma}) in SU UMa-type dwarf
novae, AB Nor was thereby strongly suspected to be an SU UMa-type dwarf
nova.  Upon this alert in VSNET, W. S. G. Walker
reported on the possible presence of a 0.4 mag superhump
(vsnet-alert 4589\footnote{
http://www.kusastro.kyoto-u.ac.jp/vsnet/Mail/alert4000/\\msg00589.html.
}).  Walker reported an approximate superhump period of 0.078--0.079 d,
based on the second-night observation
(vsnet-alert 4597\footnote{
http://www.kusastro.kyoto-u.ac.jp/vsnet/Mail/alert4000/\\msg00597.html.
}).  Although the suggested SU UMa-type nature of AB Nor was almost
confirmed by this observation, the lack of long-baseline, time-resolved
observation at this moment required us another opportunity for independent
confirmation of superhumps, as well as precisely determining their
period and evolution.

\subsection{2002 August--September Outburst}

   The next chance arrived two years later.
   The 2002 August--September Outburst was detected by Rod Stubbings
on August 31.437 UT at a visual magnitude of 14.0
(vsnet-alert 7457\footnote{
http://www.kusastro.kyoto-u.ac.jp/vsnet/Mail/alert7000/\\msg00457.html.
}).  We conducted a CCD time-series photometry campaign during this
outburst.

   The log of observation is summarized in Table \ref{tab:ablog}.

\begin{table}
\caption{Journal of the 2002 CCD photometry of AB Nor.}\label{tab:ablog}
\begin{center}
\begin{tabular}{crccrc}
\hline\hline
\multicolumn{2}{c}{2002 Date}& Start--End$^a$ & Exp(s) & $N$
        & Obs \\
\hline
Sept. &  1 & 52518.901--52519.004 & 120  &  66 & N \\
      &  1 & 52519.239--52519.366 &  45  & 173 & M \\
      &  2 & 52520.200--52520.332 &  45  & 100 & M \\
      &  4 & 52521.907--52522.008 & 120  &  64 & N \\
      & 11 & 52528.887--52528.966 & 180  &  36 & N \\
      & 12 & 52530.203--52530.343 &  50  & 159 & N \\
\hline
 \multicolumn{6}{l}{$^a$ BJD$-$2400000.} \\
\end{tabular}
\end{center}
\end{table}

\begin{figure}
  \includegraphics[angle=0,width=8.8cm]{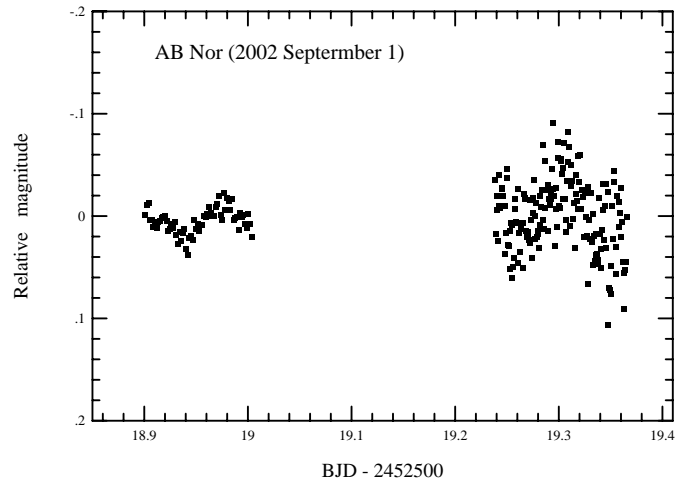}
  \caption{Light variation of AB Nor on 2003 September 1 (1 d after the
  detection of the outburst).  The amplitudes of the superhumps were
  rapidly growing.}
  \label{fig:ab1}
\end{figure}

   The initial observation was performed only one day later than the
initial detection.  The observation during this night clearly caught
the evolutionary stage of the superhumps (Figure \ref{fig:ab1}).

\begin{figure}
  \includegraphics[angle=0,width=8.8cm,height=11cm]{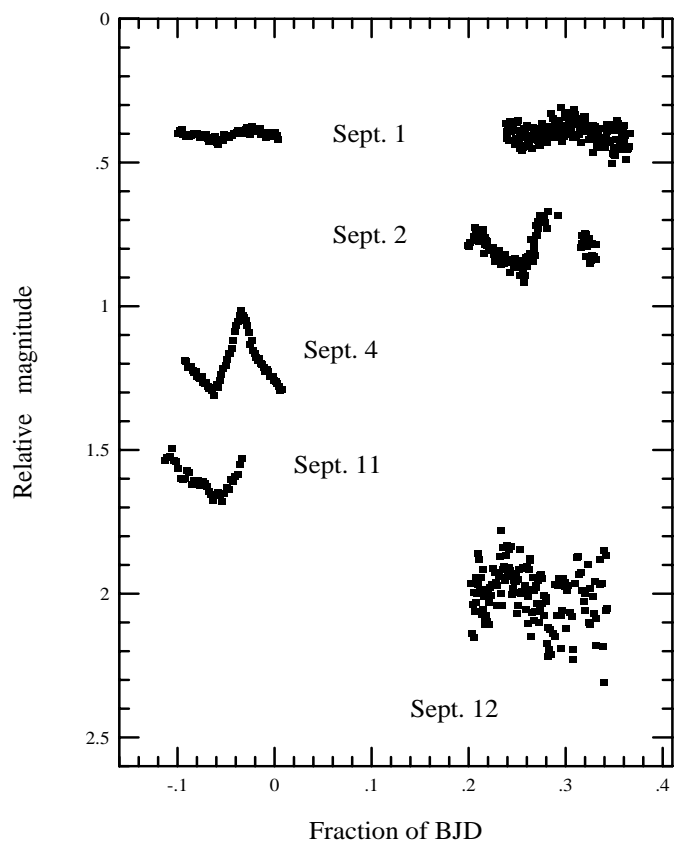}
  \caption{Nightly light curves of AB Nor.  Unavoidable gaps are
  present between observations, mainly because of the wide gap in
  longitudinal distribution of the two observers.}
  \label{fig:abday}
\end{figure}

   Figure \ref{fig:abday} shows the nightly light curves of AB Nor.
Unavoidable gaps are present between observations, mainly because of
the wide gap in longitudinal distribution of the two observers.
The superhumps had grown on September 2.  The September 12 observation
was performed just before the object started fading rapidly from
the superoutburst plateau.

\begin{figure}
  \includegraphics[angle=0,width=8.8cm]{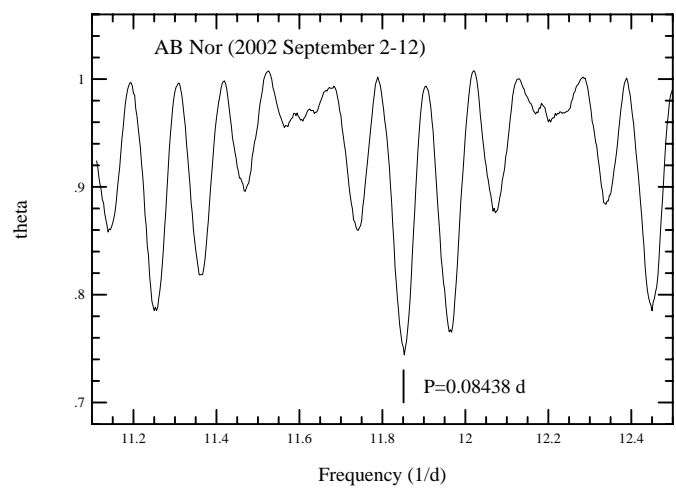}
  \caption{Period analysis of AB Nor from the September 2--12 data.}
  \label{fig:abpdma}
\end{figure}

   As is naturally expected from the early epoch observation coincident
with the rapid evolutionary stage of the superhumps, and partly owing
to the gap in observation, we have not been able to uniquely determine
the superhump period common to the entire observing period.  We thereby
divided the data into three segments (1) early evolutionary stage:
September 1, (2) fully developed stage: September 2--4,
and (3) late stage: September 11--12, and first determined the superhump
periods within respective segments.
A period analysis of the early evolutionary stage
yielded a signal around a frequency of 12.1(1) d$^{-1}$, corresponding
to a period of 0.0829(9) d.  The significance of this periodicity is
93\%.  This periodicity did not appear in the later segments, and it
likely reflected the stage of a rapid change in the superhump period.
During the latter two segments, a common frequency around 11.87(3)
d$^{-1}$, corresponding to a period of 0.0842(3) d, was present.
The selection of the alias was based on the proximity to the period
derived from the segment 1 and the common presence in segment 2
(significance $>$99.9\%) and segment 3 (significance 87\%).  From the
combination of segment 2 and 3, we obtained a period of 0.08438(2) d
(Figure \ref{fig:abpdma}).  We consider that this period is the
representative superhump period of AB Nor, although a better coverage
of a future superoutburst is desired to decisively identify the
superhump period and its evolution.

\begin{figure}
  \includegraphics[angle=0,width=8.8cm]{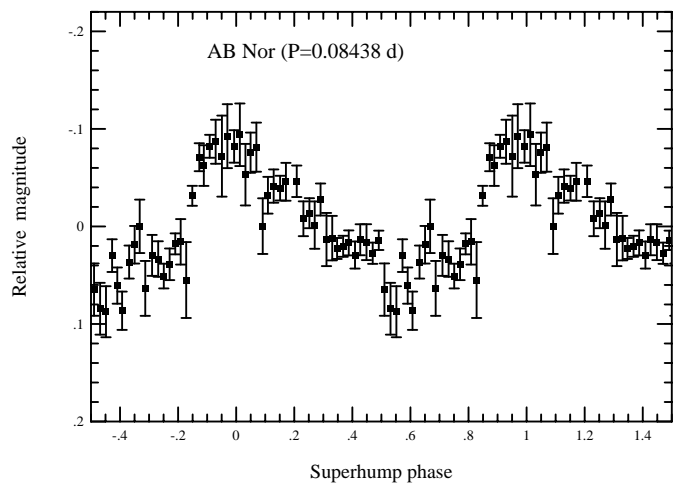}
  \caption{Mean superhump profile of AB Nor.}
  \label{fig:abph}
\end{figure}

\subsection{Astrometry and Quiescent Counterpart}

\begin{figure}
  \begin{center}
  \includegraphics[angle=0,width=8cm]{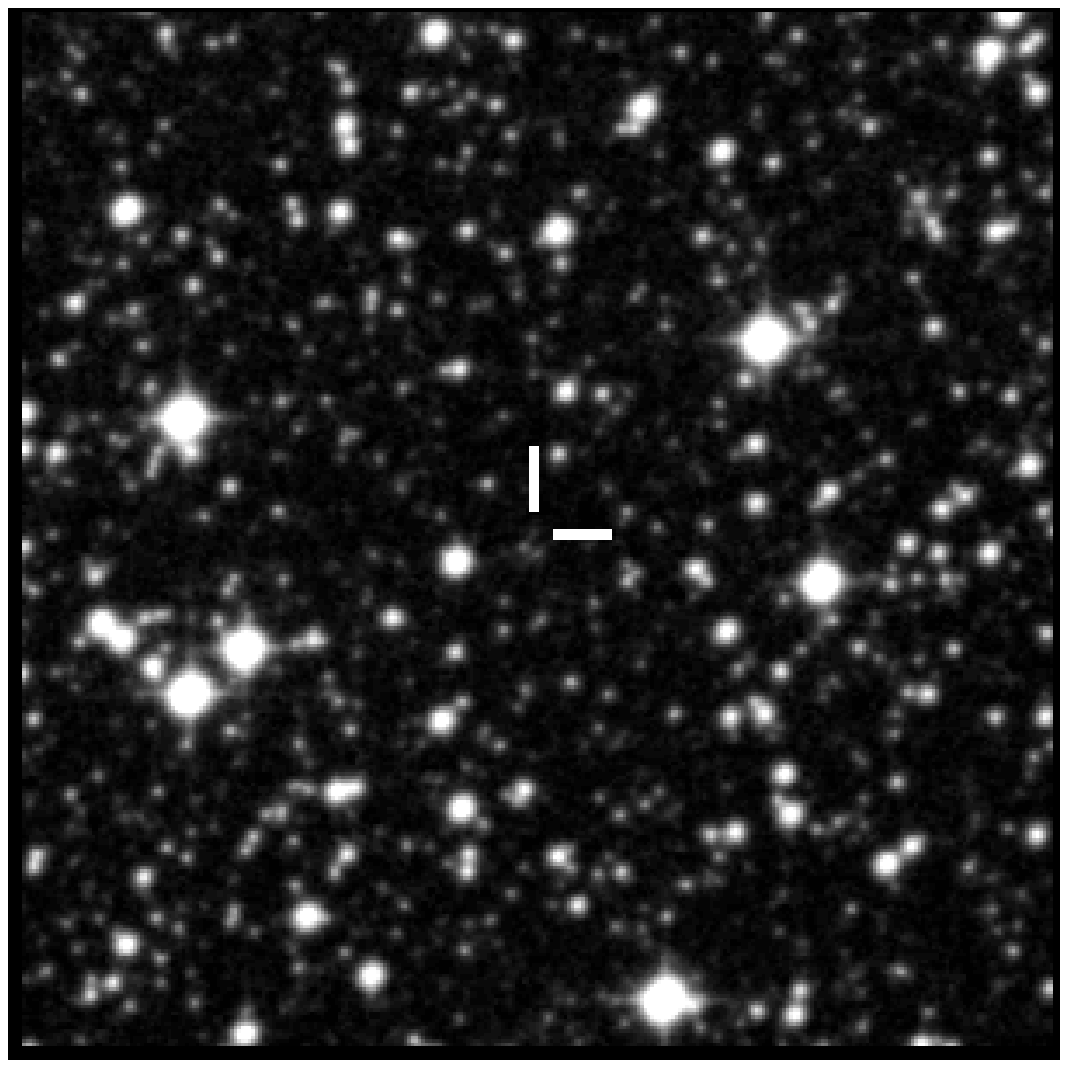} \\
  \vskip 1mm
  \includegraphics[angle=0,width=8cm]{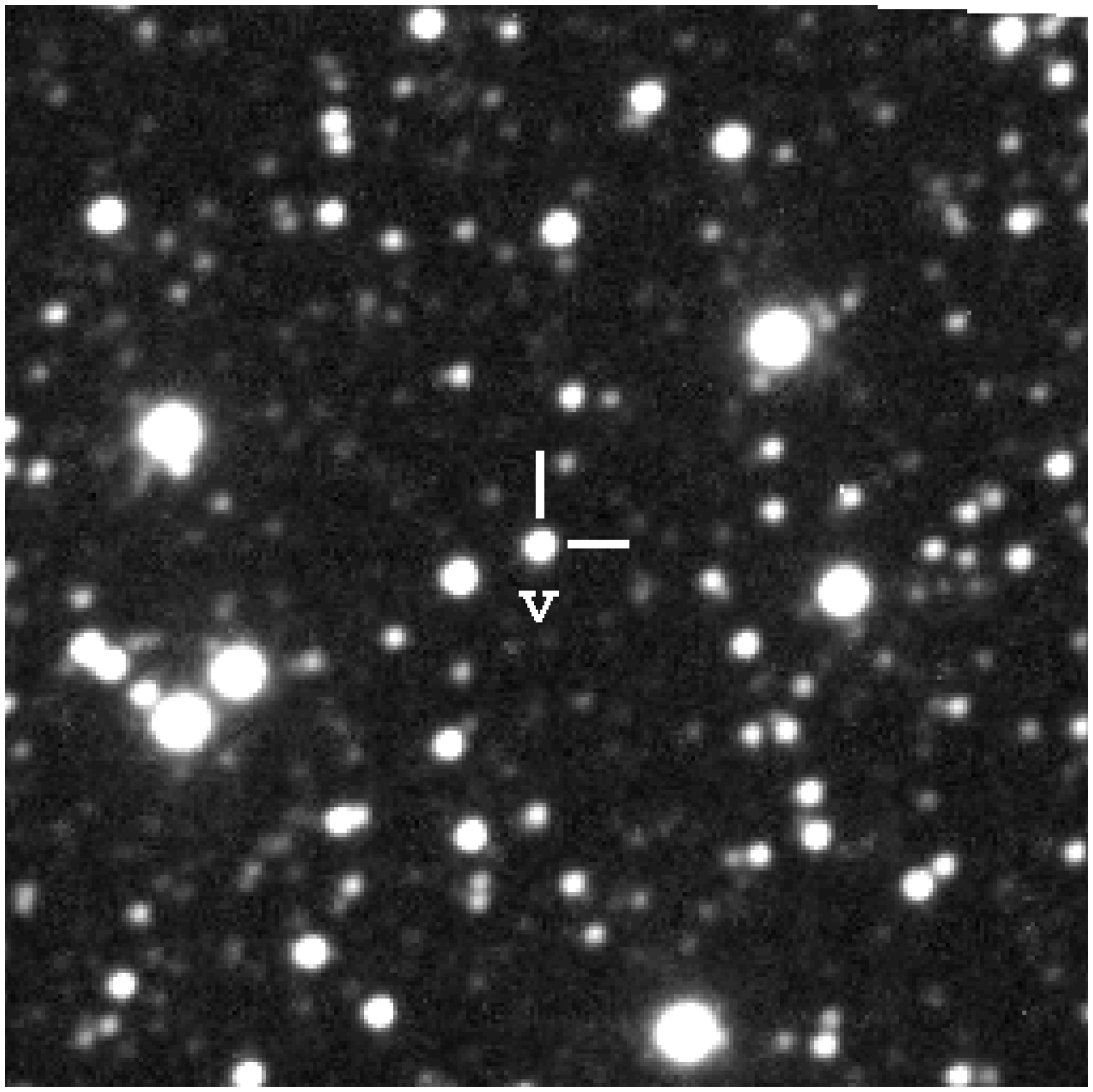}
  \end{center}
  \caption{Identification of AB Nor.  Up is north, left is east, 5
  minutes square.  (Upper) In quiescence, reproduced from the DSS 2
  red image.  (Lower) In outburst, taken on 2002 Sept. 1.438 UT by
  P. Nelson. V = AB Nor.
  }
  \label{fig:abnorid}
\end{figure}

  We measured the position of AB Nor with the outburst image taken by
P. Nelson on 2002 Sept. 1.438 UT.  The derived position with respect
to UCAC1 reference stars is
R. A. = 15$^{\rm h}$ 49$^{\rm m}$ 15$^{\rm s}$.475,
Decl. = $-$43$^\circ$ 04$'$ 48$''$.49 (J2000.0), with a fitting error of
about 0$''$.2 for each coordinate (Fig. \ref{fig:abnorid}).
This result is almost identical with the value derived by A. Henden using
2000 April outburst images taken by Walker (B. Sumner,
vsnet-chat 2800\footnote{
http://www.kusastro.kyoto-u.ac.jp/vsnet/Mail/chat2000/\\msg00800.html.
}).  The quiescent counterpart is clearly seen in every DSS images at
mag about 20.  No proper motion of this object was detected by the
examination of available archived images.

\subsection{AB Nor as an SU UMa-Type Dwarf Nova}\label{sec:abnorsu}

\begin{figure*}
  \includegraphics[angle=0,width=18cm]{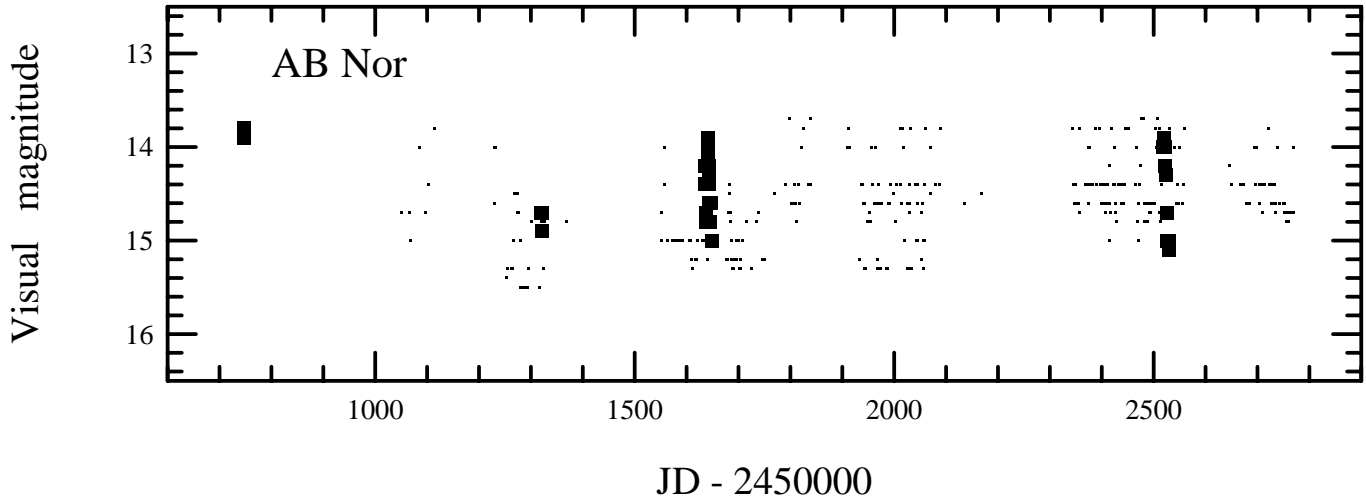}
  \caption{Long-term visual light curve of AB Nor.  Large and small dot
  represent positive detections and upper limit observations, respectively.}
  \label{fig:ablong}
\end{figure*}

   Figure \ref{fig:ablong} shows the long-term visual light curve of AB Nor.
Table \ref{tab:about} lists the observed outbursts.
Two well-defined superoutbursts (JD 2451636 and 2452517)
with durations longer
than 12 d are unambiguously identified.  The outburst on JD 2450746
is also likely a superoutburst based on its brightness.  The outburst
on JD 2451320 is probably a normal outburst based on its faintness.
Although there are observational gaps, there seems to be little chance
of many missed outbursts.  The supercycle of AB Nor is thereby
estimated to be $\sim$880/$N$ d, where $N$ is either 1 or 2.  We suggest
that AB Nor belongs to SU UMa-type dwarf novae with long supercycle
lengths.

\begin{table}
\caption{List of Outbursts of AB Nor.}\label{tab:about}
\begin{center}
\begin{tabular}{ccccc}
\hline\hline
JD start$^a$ & JD end$^a$ & Max & Duration (d) & Type \\
\hline
50746.9 & 50747.9 & 13.8 & $>2$ & super? \\
51320.3 & 51322.2 & 14.7 & 2    & normal \\
51636.3 & 51649.2 & 14.3 & 13   & super \\
52517.9 & 52529.9 & 14.0 & 12   & super \\
\hline
 \multicolumn{5}{l}{$^a$ JD$-$2400000.} \\
\end{tabular}
\end{center}
\end{table}

   The derived superhump period of 0.08438 d is the one of the longest
periods among SU UMa-type dwarf novae below the period gap.
The other long-period systems include TY PsA ($P$=0.08765 d:
\citealt{bar82typsa,war89typsa}), BF Ara ($P$=0.08797 d: \citealt{kat03bfara})
and YZ Cnc ($P$=0.09204 d: \citealt{pat79SH,vanpar94suumayzcnc,kat01yzcnc}),
which show frequent outbursts and superoutbursts.  Among SU UMa-type
dwarf novae with similar superhump periods, EF Peg ($P$=0.08705 d:
\citealt{kat02efpeg}), V725 Aql ($P$=0.09909 d: \citealt{uem01v725aql})
and DV UMa ($P$=0.08869 d: \citealt{nog01dvuma})
have a low frequency of outbursts comparable to that of AB Nor.
Both EF Peg and V725 Aql are considered to be unusual in its outburst
frequency and behavior \citep{uem01v725aql,kat02efpeg}, and may have low
mass-transfer rate comparable to WZ Sge-type dwarf novae
\citep{kat01hvvir}.  Being easily observable at minimum (compared to
EF Peg and V725 Aql), further detailed observation of AB Nor in quiescence
will be helpful identifying the nature of these long-period SU UMa-type
dwarf novae with supposed low mass-transfer rates.

\section{CAL 86}

\subsection{Introduction}

   CAL 86 = 1RXP J054610$-$6835.1 = 1RXS J054613.6$-$683523 is
a cataclysmic variable in the direction of
the Large Megellanic Cloud (LMC).  This star was
originally selected as an {\it Einstein} X-ray source.
\citet{sch02cal86} reported the detection of its short (0.066 d)
orbital period and at least five outbursts from the MACHO observations.
Some of the outbursts reached $V$ = 14 (amplitude 5 mag).
This orbital period, together with the presence of large-amplitude
outbursts, makes CAL 86 a good SU UMa-type candidate.
Upon this information we undertook a monitoring campaign
(vsnet-campaign-dn 2561\footnote{
http://www.kusastro.kyoto-u.ac.jp/vsnet/Mail/\\campaign-dn2000/msg00561.html.
}) since 2002 December.  Only one outburst was observed (in 2003 February)
up to 2003 August.

\subsection{2003 February Outburst}

\begin{figure}
  \includegraphics[angle=0,width=8.8cm]{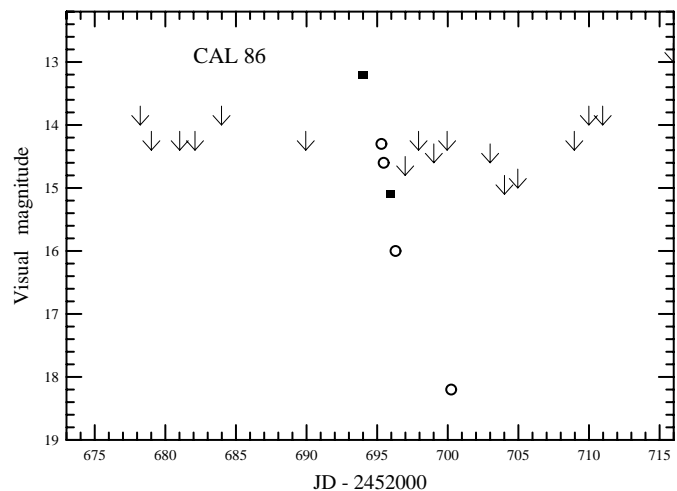}
  \caption{The 2003 February outburst of CAL 86.  The filled squares
  and downward arrows represent positive detections and upper limit
  observations, respectively.
  The open circles represent Monard's snapshot unfiltered CCD photometry.}
  \label{fig:calvis}
\end{figure}

   The 2003 February outburst was detected at a visual magnitude of
13.2 on February 23.454 UT by Rod Stubbings (vsnet-alert 7645\footnote{
http://www.kusastro.kyoto-u.ac.jp/vsnet/Mail/alert7000/\\msg00645.html.
}).  The outburst very quickly faded after the outburst detection
(Figure \ref{fig:calvis}).  The mean fading rate of the initial 2.5 d
was 1.1 mag d$^{-1}$, which is a typical value for a normal outburst
of an SU UMa-type dwarf nova \citep{BaileyRelation,kat02gycnc}.

   The log of observation is summarized in Table \ref{tab:cal86log}.

\begin{table}
\caption{Journal of the 2003 CCD photometry of CAL 86.}\label{tab:cal86log}
\begin{center}
\begin{tabular}{crccrc}
\hline\hline
\multicolumn{2}{c}{2003 Date}& Start--End$^a$ & Exp(s) & $N$
        & Obs \\
\hline
Feb.  & 24 & 52695.228--52695.467 &  45  & 315 & M \\
      & 25 & 52696.038--52696.187 &  45  & 251 & H \\
\hline
 \multicolumn{6}{l}{$^a$ BJD$-$2400000.} \\
\end{tabular}
\end{center}
\end{table}

\begin{figure}
  \includegraphics[angle=0,width=8.8cm]{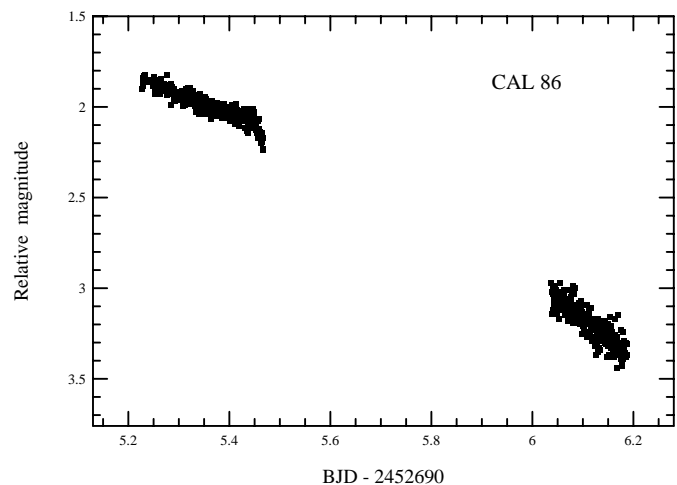}
  \caption{The 2003 February outburst of CAL 86 drawn from the time-resolved
  CCD observations.  The magnitudes are given relative to GSC 9163.607
  (approximate $R_{\rm c}$ magnitude 12.4).  A rapid, smooth decline is
  apparent.}
  \label{fig:callc}
\end{figure}

\begin{figure}
  \includegraphics[angle=0,width=8.8cm]{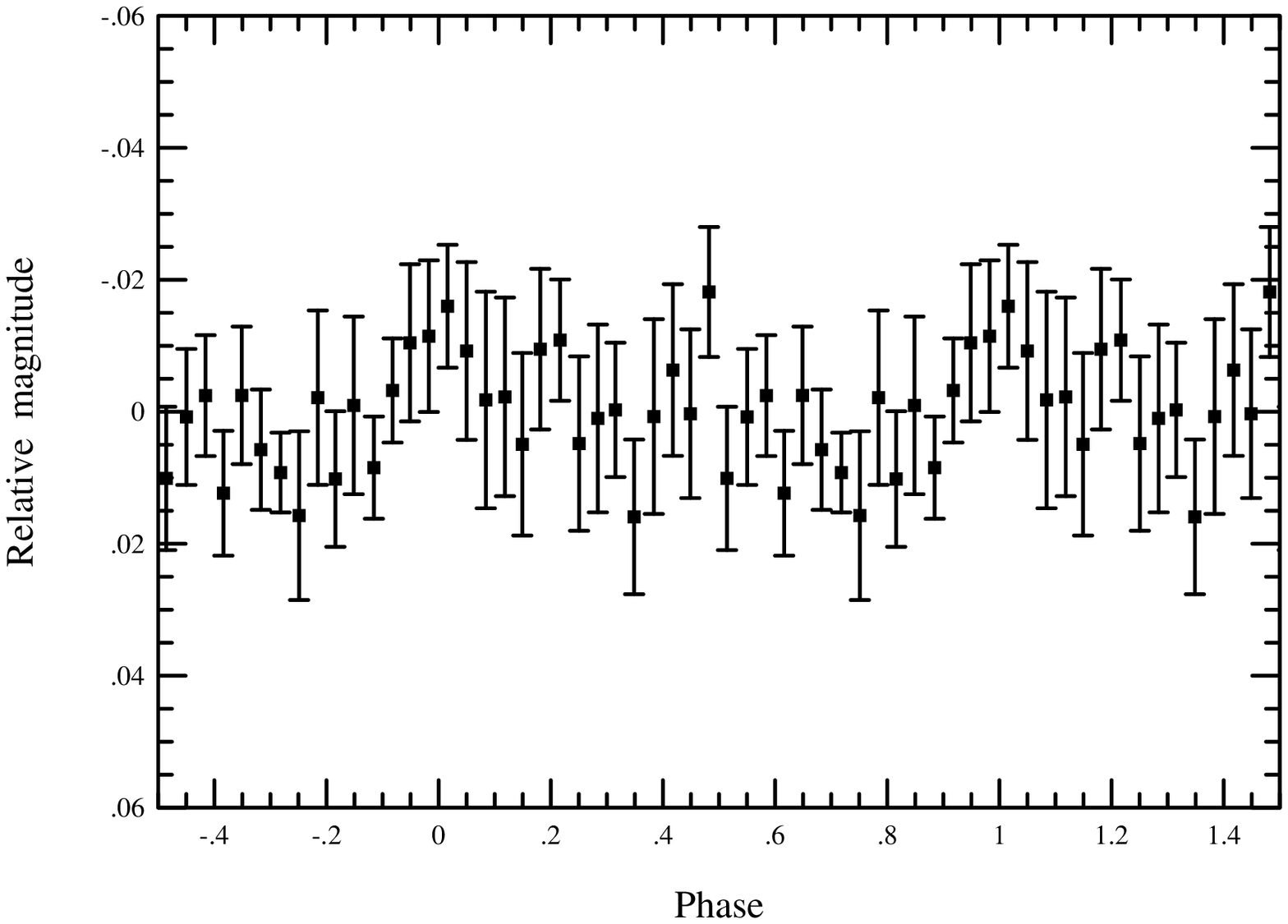}
  \caption{``Orbital" light curve phase-averaged at the reported orbital
  period of 0.066 d.  Only a marginal (0.02 mag) modulation was detected.}
  \label{fig:calph}
\end{figure}

   Figure \ref{fig:callc} shows the light curve drawn from the time-resolved
CCD observations.  The object was rapidly and smoothly fading during this
observing period.  A period analysis of the data did not reveal any
superhump-type variation with an amplitude larger than 0.05 mag.
Figure \ref{fig:calph} shows an ``orbital" light curve phase-averaged
at the reported orbital period of 0.066 d, after removing the trend
of steady decline from Figure \ref{fig:callc}.  Only a marginal (0.02 mag)
modulation was detected, which is not inconsistent with the general
lack of orbital signatures in outbursting non-eclipsing dwarf novae
(see also \citealt{kat01aqeri}).

\subsection{Astrometry and Quiescent Counterpart}

\begin{figure}
  \begin{center}
  \includegraphics[angle=0,width=8cm]{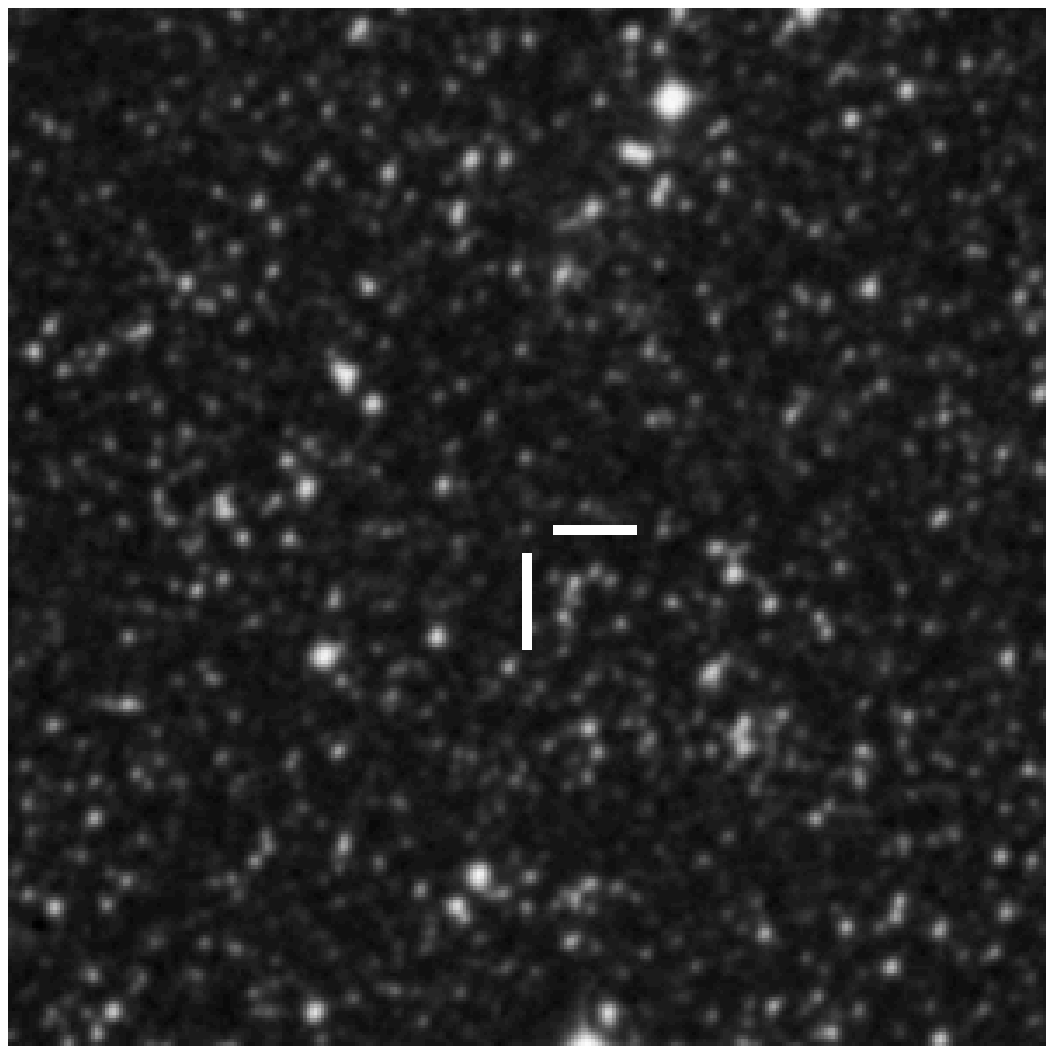} \\
  \vskip 1mm
  \includegraphics[angle=0,width=8cm]{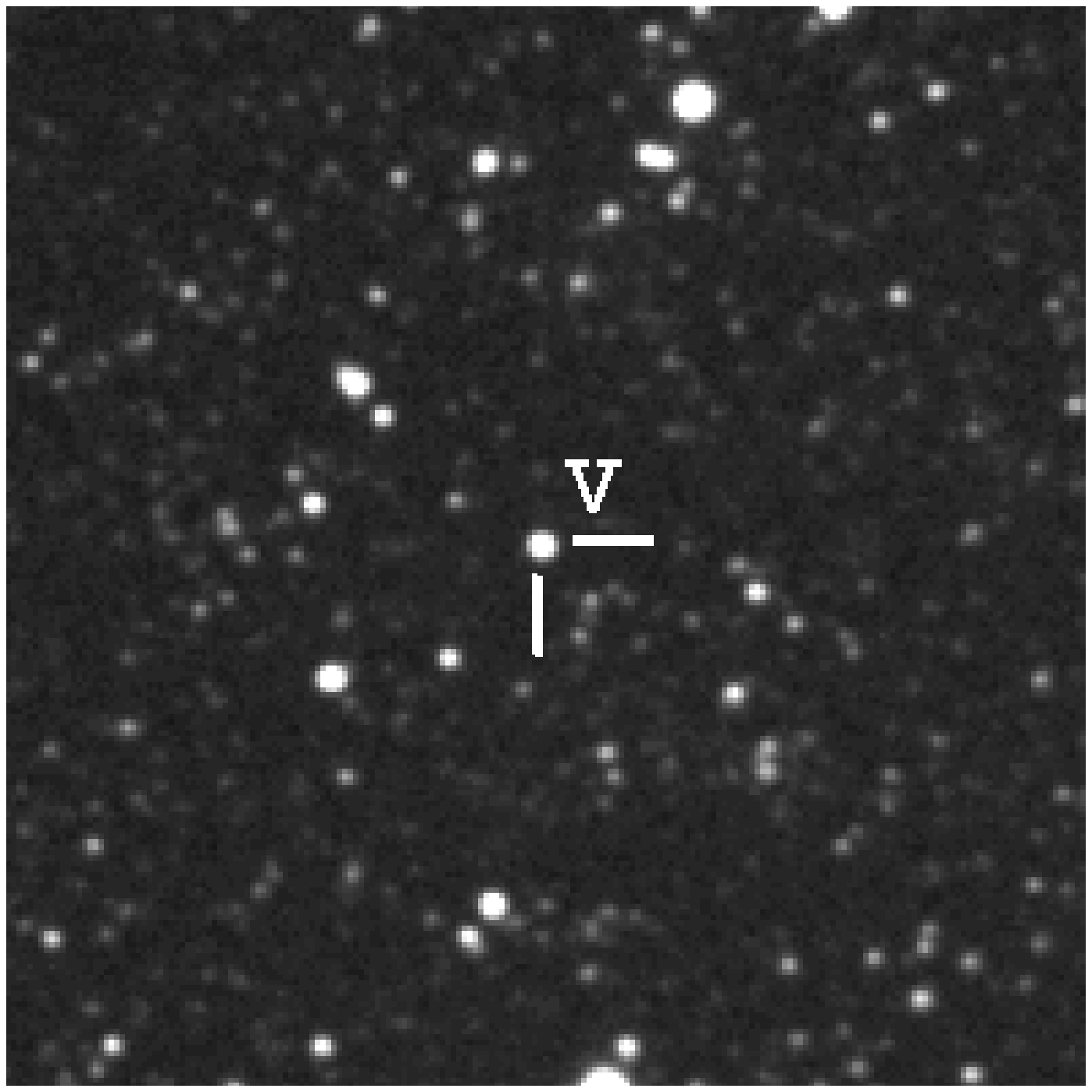}
  \end{center}
  \caption{Identification of CAL 86.  Up is north, left is east, 5
  minutes square.  (Upper) In quiescence, reproduced from the DSS 2
  red image.  (Lower) In outburst, taken on 2003 Feb. 24.79 by
  B. Monard.  V = CAL 86.
  }
  \label{fig:cal86id}
\end{figure}

   Since CAL 86 is located in the LMC field with a huge number of faint
stars, we tried to make independent astrometry and identification using
the outburst CCD images.  The position with respect to UCAC1 frame was
derived to be R. A. = 05$^{\rm h}$ 46$^{\rm m}$ 14$^{\rm s}$.973, Decl. = 
$-$68$^\circ$ 35$'$ 23$''$.76 (J2000.0), with a fitting error less
than 0$''$.1 for each coordinate (Fig. \ref{fig:cal86id}).
This star is identical to the one labeled as ``Star No. 2" in the chart
of \citet{sch94LMCSSS}, and to a USNO-B1.0 star having position end
figures of 14$^{\rm s}$.98, 24$''$.1 ($r_2$ mag 18.77).  The examination
of archived images revealed no detectable proper motion of this
object, which was accidentally caught in outburst on an image taken
on 1987 Jan. 24 as noted in Downes et al.'s online catalog.
The USNO-B1.0 entry also shows no proper motion.

\section{Summary}

We photometrically observed four southern dwarf novae in outburst
(NSV 10934, MM Sco, AB Nor and CAL 86).  We succeeded in measuring
the superhump periods of the first three systems, and clarified the
long-term outburst characteristics from long-term visual observations.

(1) NSV 10934 was confirmed to be
an SU UMa-type dwarf nova with a mean superhump period of
0.07478(1) d.  The star also showed transient appearance of quasi-periodic
oscillations (QPOs) during the final growing stage of the superhumps.
Combined with the recent theoretical interpretation and with the rather
unusual rapid terminal fading of normal outbursts, NSV 10934 may be
a candidate intermediate polar showing SU UMa-type properties.

(2) We determined the mean mean superhump periods of the newly identified
SU UMa-type dwarf nova MM Sco to be 0.06136(4) d.  The combination of
a short superhump period and a low frequency of outbursts suggests
that MM Sco belongs to a class of infrequently outbursting SU UMa-type
dwarf novae resembling UV Per.  The true quiescence of MM Sco may
be fainter than has been believed.

(3) We determined the mean superhump period of AB Nor, whose SU UMa-type
nature is established by this study, to be 0.08438(2) d.
We suggest that AB Nor belongs to a rather rare class of
long-period SU UMa-type dwarf novae with low mass-transfer rates.

(4) We also observed an outburst of the
suspected SU UMa-type dwarf nova CAL 86.  We identified this outburst
as a normal outburst and determined the mean decline rate of
1.1 mag d$^{-1}$.

\section*{Acknowledgments}

This work is partly supported by a grant-in-aid [13640239, 15037205 (TK),
14740131 (HY)] from the Japanese Ministry of Education, Culture, Sports,
Science and Technology.
The CCD operation of the Bronberg Observatory is partly sponsored by
the Center for Backyard Astrophysics.
The CCD operation by Peter Nelson is on loan from the AAVSO,
funded by the Curry Foundation.
This research has made use of the Digitized Sky Survey producted by STScI, 
the ESO Skycat tool, the VizieR catalogue access tool.

\label{lastpage}

\end{document}